\documentclass[twocolumn,aps,prl,superscriptaddress]{revtex4-1}
\usepackage{here}
\usepackage{graphicx}
\usepackage{braket}
\usepackage{color}
\usepackage{amsmath}

\makeatother

\begin{document}

\title{Magnetic Field Inducing Zeeman Splitting and Anomalous Conductance
Reduction of Half-integer Quantized Plateaus in InAs Quantum Wires}

\author{Sadashige Matsuo}

\email{matsuo@ap.t.u-tokyo.ac.jp}

\affiliation{Department of Applied Physics, University of Tokyo, Bunkyo-ku, Tokyo
113-8656, Japan}

\author{Hiroshi Kamata}

\affiliation{Department of Applied Physics, University of Tokyo, Bunkyo-ku, Tokyo
113-8656, Japan}
\affiliation{Center for Emergent Materials Science, RIKEN, Wako, Saitama, Japan}

\author{Shoji Baba}

\affiliation{Department of Applied Physics, University of Tokyo, Bunkyo-ku, Tokyo
113-8656, Japan}

\author{Russell S. Deacon}

\affiliation{Center for Emergent Materials Science, RIKEN, Wako, Saitama, Japan}
\affiliation{Advanced Device Laboratory, RIKEN, Wako, Saitama, Japan}

\author{Javad Shabani}

\affiliation{California NanoSystems Institute, University of California, Santa
Barbara, CA 93106, USA}

\author{Christopher J. Palmstr\o m}

\affiliation{California NanoSystems Institute, University of California, Santa
Barbara, CA 93106, USA}
\affiliation{Electrical and Computer Engineering, University of California, Santa 
Barbara, CA 93106 USA}
\affiliation{Materials Department, University of California, Santa
Barbara, CA 93106, USA}

\author{Seigo Tarucha}
\email{tarucha@ap.t.u-tokyo.ac.jp}

\affiliation{Department of Applied Physics, University of Tokyo, Bunkyo-ku, Tokyo
113-8656, Japan}
\affiliation{Center for Emergent Materials Science, RIKEN, Wako, Saitama, Japan}

\begin{abstract}
We report on magnetic field dependence of half-integer quantized conductance plateaus (HQPs) in InAs quantum wires. We observed HQPs at zero applied magnetic field in InAs quantum wires fabricated from a high-quality InAs quantum well. The application of in-plane magnetic field causes Zeeman splitting of the HQP features, indicating that the origin of the observed HQP is not spontaneous spin polarization. Additionally we observe that conductance of the split HQPs decreases gradually as the in-plane magnetic field increases. We finally assume electron-electron interaction as a possible mechanism to account for the zero-field HQPs and the anomalous field dependence. 
\end{abstract}
\maketitle
Since the seminal discovery of quantized conductance plateaus in quantum wires~\cite{weesprl1988}, anomalous conductance plateau quantization which is widely reported experimentally has been an intriguing topic in mesoscopic physics. This topic has been studied extensively from various aspects including Tomonaga-Luttinger liquid (TLL), spontaneous spin polarization, Kondo effect. In 1990's, it was reported that quantized plateau conductance of long GaAs quantum wires decreases at high temperature but maintains the plateau shape observed at lower temperatures~\cite{taruchassc1995, yacobyprl1996, levyprl2006}. This conductance reduction was explained as a property of the TLL, a one-dimensional electron liquid with electron-electron (e-e) interaction~\cite{tomonagaptp1950, kaneprb1992, ogataprl1994, maslovprb1995}. In 1996 another anomalous feature, the so-called 0.7 anomaly, was reported in GaAs quantum wires~\cite{thomasprl1996}. The anomaly appears as a shoulder or plateau-like structure at $0.7\times2e^{2}/h$ in addition to the quantized conductance plateaus. Although the 0.7 anomaly has been theoretically and experimentally studied ~\cite{thomasprb1998,kristensenprb2000,reillyprb2001,cronenwettprl2002,rokhinsonprl2006,dicarloprl2006,bauernature2013,iqbalnature2013}, the origin remains a subject of debate. 

Furthermore, anomalous conductance plateaus at half-integer multiples of quantized conductance (HQP) have been observed even at zero magnetic field ($B=0$ T) in GaAs quantum wires~\cite{crookscience2006, hewpr;2008}, carbon nanotubes~\cite{biercukprl2005}, and InAs (InGaAs) quantum wires~\cite{debraynatnano2009, wanprb2009, kohdanatcommun2012}.  The observation of HQPs at $B=0$  T is striking because we naively expect that spin states are degenerate, resulting in plateaus at multiples of $2e^{2}/h$, and therefore that steps quantized at $e^{2}/h$ appear only in a spin-resolved quantum wire at finite magnetic field. HQPs were often reported in quantum wires with intrinsic or electric-field induced spin-orbit interaction (SOI). Therefore, the origin is sometimes interpreted as spin-related phenomena caused by SOI. Although there are several theoretical suggestions for the origin, such as spontaneous spin polarization~\cite{debraynatnano2009, wanprb2009,ngoprb2010}, Stern-Gerlach mechanism~\cite{kohdanatcommun2012} and "spin-incoherent" Luttinger liquid (SILL)~\cite{matveevprb2004, fietermp2007}, the underlying physics is still controversial, much like the 0.7 anomaly. According to the spontaneous spin polarization (SSP) scenario, the lateral SOI and the e-e interaction invokes SSP resulting in the apperance of zero-field HQPs. Stern-Gerlach mechanism (SG) suggests that the spin-filter effect invoked by SOI around entrance of the wire results in different transmission probabilities, 0 and 1, for spin-up and down electrons, respectively, resulting in the zero-field HQPs. Unlike the above two scenarios, SILL does not require SOI but relies on a finite temperature effect preventing the spin density wave mode from propagating.  In experiments to feature the zero field HQPs, the in-plane magnetic field dependence is critical because SSP and SG scenarios predict peculiar magnetic field dependence of HQPs that is not expected in conventional quantized conductance plateaus. The field dependence can also have important implications in recent studies on the spin effects in hybrid semiconductors-superconductor devices. Indeed application of in-plane magnetic field is a key ingredient for the creation of Majorana Fermions in a combined system of a quantum wire with strong SOI and a s-wave superconductor~\cite{fuprl2008,hasanrmp2010,mourikscience2012,dasnatphys2012,  rokhinsonnatphys2012,albrechtnat2016}.

Here we report on an experimental study of in-plane magnetic field dependence of the zero field HQPs observed in InAs quantum wires fabricated from a quantum well. We discuss validity of the most likely scenarios for the observed HQPs including SSP, SG, and SILL but not all possible mechanisms. Note we here study in detail Zeeman effect because SSP, SG, and SILL predict different characteristic dependence on $B$ allowing the valid mechanism for HQPs to be identified.

Since InAs has a large g-factor, resulting in large Zeeman effect and our quantum well has high mobility even at low carrier density, we clearly observe the Zeeman splitting of HQP with an in-plane magnetic field, indicating that SSP does not occur. Observed magnetic field dependence is symmetric about the $B=0$ T, indicating that SG is not the mechanism at play. Furthermore we found that increase of the in-plane magnetic field makes the conductance of HQPs smaller. We propose that these observed phenomena may be assigned to SILL in the quantum wire. 
 
\begin{figure}[t]
\centering \includegraphics[width=0.9\linewidth]{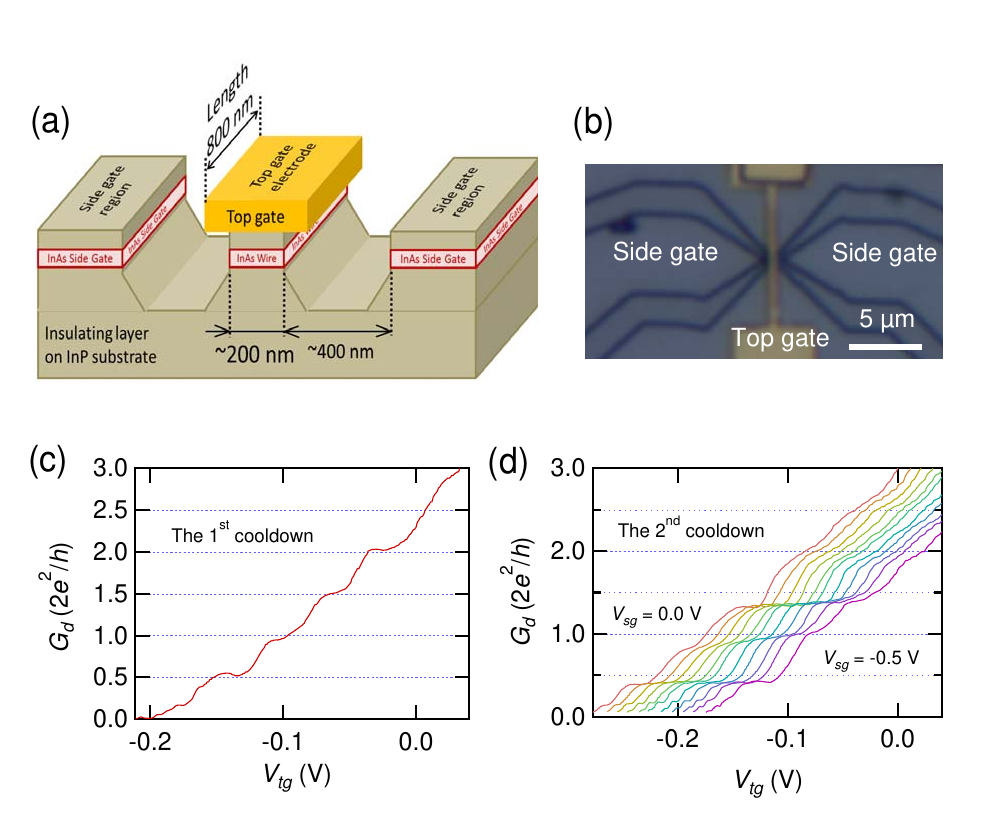} 
\caption{(a)A schematic cross section of the fabricated quantum wire devices. (b)An optical image of the 0.8 ${\rm \mu}$m-long wire device. The yellow part is a top gate electrode and there are two side gate electrodes. (c)The differential conductance, $G_d$ as a function of top gate voltage, $V_{tg}$ obtained at $T=1.5$ K in the 1st cooldown of the device is shown. There are 4 conductance plateaus observed at 0.5, 1.0, 1.5, and 2.0 $\times 2e^2/h$. (d)$G_d$ vs. $V_{tg}$ in the 2nd cooldown of the device. Each curve is obtained at different side gate voltage, $V_{sg}$ between $-0.5~{\rm V}<V_{sg}<0.0~{\rm V}$. The conductance of plateaus has a negligible dependence on $V_{sg}$g.}
\label{fig1} 
\end{figure}
 We fabricated quantum wire devices from a two dimensional electron gas (2DEG) in InAs quantum wells~\cite{shabaniprb2014,shabaniapl2014}. The InAs well is 4 nm thick and the carrier density and mobility of the 2DEG are $3\times 10^{11}~{\rm cm^{-2}}$ and $10~{\rm m^2V^{-1}s^{-1}}$, respectively. The mobility is remarkably high for such a low 2DEG density as compared to that used in most of the previous reports on HQP. The e-e interaction plays an important role in the electron transport of quantum wires with a low carrier density, because the interaction strength is proportional to the carrier density $n$ while the kinetic energy is proportional to $n^2$. 

A schematic cross section of the fabricated quantum wire is shown in Fig.~\ref{fig1}(a). In order to form a 200 nm-wide quantum wire with side gate electrodes, the unnecessary parts of the 2DEG were etched away using a H$_{3}$PO$_{4}$ based etchant following the fabrication of a top-gate electrode (Ti 5 nm Au 30 nm) which is used as a mask. In this report we focus on a wire of 0.8 ${\rm \mu}$m length but to investigate the reproducibility, we have also measured a 1.4 ${\rm \mu}$m-long wire. Figure~\ref{fig1}(b) shows an optical microscope picture of the 0.8 ${\rm \mu}$m-long wire. The calculated mean free path of the 2DEG is about 2.7 ${\rm \mu}$m, which is sufficiently larger than the wire length. This means the wire transport is ballistic. We measure the differential conductance, $G_d=dI_d/dV_{sd}$ where $I_d$, and $V_{sd}$ are the drain current, and source-drain voltage, respectively using a four terminal method to eliminate contributions from the conductance of the 2DEG in regions other than the wire. Measurements are performed in the temperature range of $T=3.4$ K to $1.5$ K using standard lock-in techniques. Due to restrictions of our setup, we warmed up the devices once when changing the orientation of the in-plane magnetic field. We applied in-plane magnetic fields parallel and perpendicular to the wire in the 1st and 2nd cool down, respectively. 

Figure~\ref{fig1}(c) shows $G_d$ as a function of the top gate voltage, $V_{tg}$ at a side gate voltage of $V_{sg}=-0.5$ V at $T=$1.5 K in the 1st cooldown of the device. We applied the same negative gate voltage on both side gate electrodes to suppress the conduction from possible surface accumulation states. In this figure we observe clear conductance plateaus at the half-quantized conductance, $G_d=0.5~{\rm and}~1.5\times2e^{2}/h$. In contrast to previous reports, in which zero field HQP only appear at $G_d=0.5\times2e^{2}/h$, we observe an additional plateau at $G_d=1.5\times2e^{2}/h$. We note that the wire was unstable when sweeping $V_{sg}$, especially below -0.5 V probably due to charging at the etched surface. Therefore, we fixed $V_{sg}$ and varied the carrier density using $V_{tg}$. In the 2nd cooldown, we observe similar HQPs in the range of -0.5 V$\leq V_{sg}\leq0$V as shown in Fig.~\ref{fig1}(d). We note that HQPs are observed even at $V_{sg}=$0 V. To characterize the wire properties, we measured temperature dependence and bias voltage, $V_{sd}$ dependence, successfully demonstrating that the device shows the expected features of a one-dimensional electron system, excluding the anomalous plateau conductance. In these measurements, no evidence of the 0.7 anomaly and no apparent signatures from impurities are found (see the SM).

\begin{figure}[t]
\centering \includegraphics[width=0.9\linewidth]{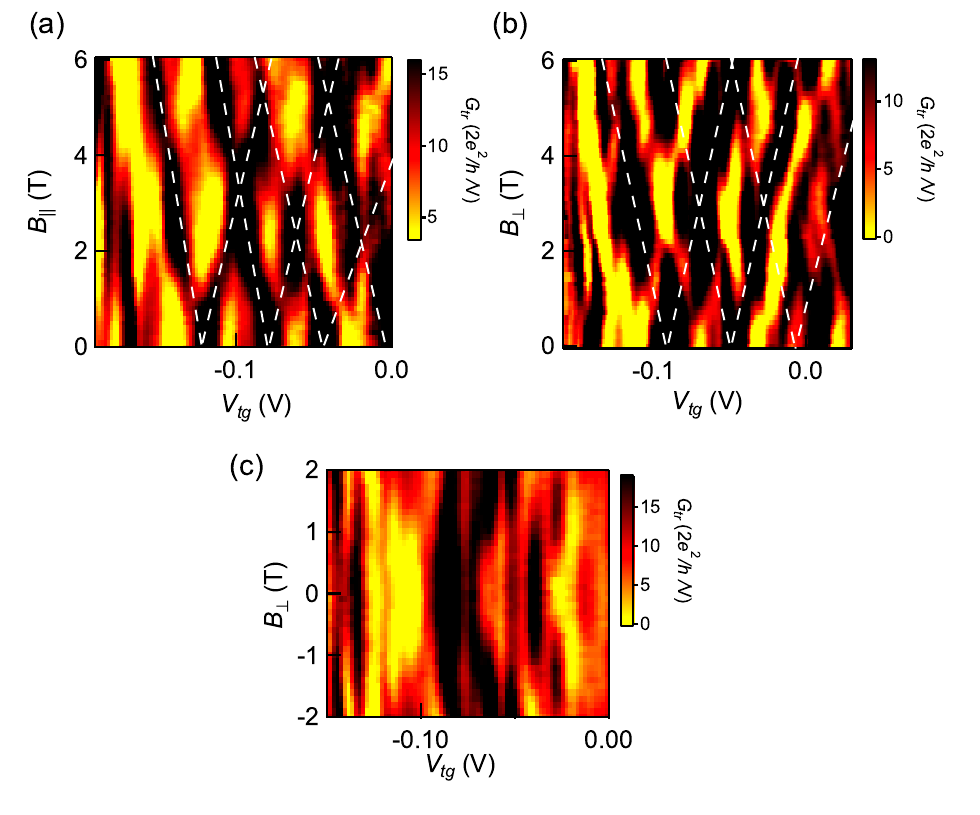} 
\caption{(a)$G_{tr}$ as functions of $V_{tg}$ and $B_{\parallel}$. Zeeman splitting of the subbands is clearly seen as diamond-shaped structures, indicated with white dash lines. (b) $G_{tr}$ as functions of $V_{tg}$ and $B_{\perp}$. The field dependence is very similar to the dependence in the $B_{\parallel}$ case. (c) $G_{tr}$ as functions of $V_{tg}$ and $B_{\perp}$ around $B_{\perp}$=0 T. The observed structures are symmetric about $B_{\perp}$=0 T.}
\label{fig2} 
\end{figure}
Now we investigate the in-plane magnetic field dependence of the trans-conductance, $G_{tr}$ to study the mechanism generating zero field HQPs, specifically whether the HQPs are related to the spin-related phenomena, SSP and SG. First, we applied an in-plane magnetic field parallel to the wire, $B_{\parallel}$. $G_{tr}$ as functions of $V_{tg}$ and $B_{\parallel}$ is plotted in Fig.~\ref{fig2}(a). The bright regions (indicating low $G_{tr}$) correspond to the conductance plateaus, while the dark regions indicate the plateau-transitions forming diamond-like features with white dash lines as a guide for the eye. In this case, the diamond-like feature is explained by the Zeeman effect resolving the spin degeneracy as previously observed in p-typed GaAs quantum wires under magnetic fields~\cite{komijaniepl2013, fabrizioprl2014}. The Zeeman effect makes the dark regions split into two dark lines as the magnetic field increases from $B_{\parallel}$=0 T. We can convert $V_{tg}$ to an energy using the results of bias measurement (see the SM) allowing the evaluation of the g-factor from the splitting of the dark lines, resulting in 5.1 for the $0.5\times2e^{2}/h$ plateau. This value is consistent with that previously reported for InAs systems, (quantum dot; 3 $\sim$  9, bulk; 14) [37-40]. The present observation clearly indicates that the subbands are spin-degenerate at $B_{\parallel}$=0 T and the degeneracy is lifted by application of $B_{\parallel}$.

To further confirm the Zeeman effect, we measured $G_{tr}$ in the presence of an in-plane magnetic field $B_{\perp}$ perpendicular to the wire, and plot the obtained result in Fig.~\ref{fig2}(b). As expected from the $B_{\parallel}$ experiment of Fig.~\ref{fig2}(a), we observe a diamond-shaped $G_{tr}$ pattern produced by the Zeeman effect. The g factor derived from the white dash lines which split from transition between the 0.5 and 1 $\times 2e^2/h$ plateaus at 0 T is 5.9, consistent with that obtained from the $B_{\parallel}$ dependence. There is no qualitative difference between the $B_{\perp}$ and $B_{\parallel}$ measurements as shown in Fig.~\ref{fig2}(a) and (b). In addition, we closely measured the $B_{\perp}$ dependence of the $G_{tr}$ vs. $V_{tg}$ around $B_{\perp}$= 0 T shown in Fig.~\ref{fig2}(c). All of the main features in Fig.~\ref{fig2}(c) are symmetric about $B_{\perp}$= 0 T.

\begin{figure}[t]
\centering \includegraphics[width=0.9\linewidth]{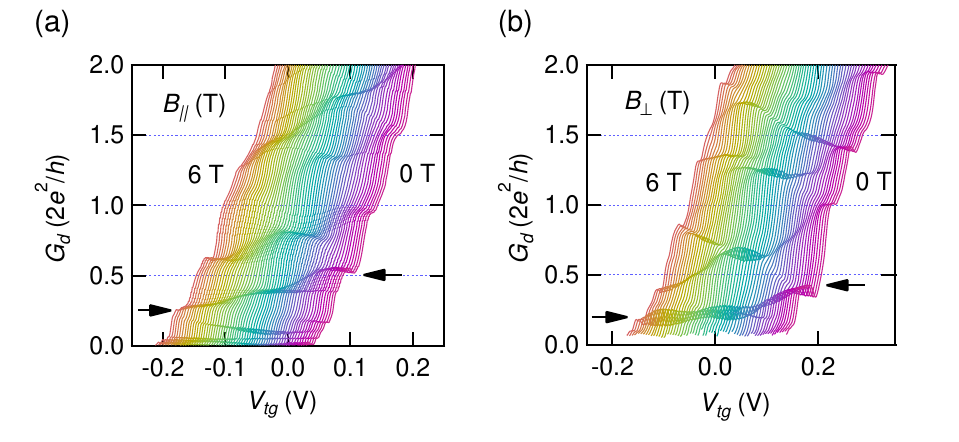}
\caption{(a) $G_d$ vs. $V_{tg}$ obtained at different $B_{\parallel}$ are shown. The horizontal axis shows the result at 6 T and the other results are incrementally shifted by 0.004 V for clarity. The lowest plateau conductance is less than 0.5$\times 2e^2/h$ and the plateau conductance gradually decreases as $B_{\parallel}$ increases. (b)$G_d$ vs. $V_{tg}$ obtained at different $B_{\perp}$ are shown. The horizontal axis shows the result at $B_{\perp}$= 6 T and the other results are incrementally shifted by 0.005 V for clarity. The main features are the same as obtained in the $B_{\parallel}$ case. }
\label{fig3} 
\end{figure}
We now consider the plateau conductance in finite magnetic fields. Figure~\ref{fig3}(a) shows $G_d$ at $V_{sd}=0$ V as a function of $V_{tg}$ measured for various values of $B_{\parallel}<$6 T in (c) and $B_{\perp}<$6 T in (d). Surprisingly the conductance of the lowest plateau at high magnetic fields is smaller than $e^{2}/h$, as highlighted by arrows on the $B=6$ T traces in both figures. The conductance gradually decreases, starting from a value of about $e^{2}/h$ (indicated by arrows on the $B=0$ T traces) as the magnetic field increases. These anomalous magnetic field dependences are obtained for the conductance measured at $V_{sd}=0$ V, so the mechanism is different from the bias-induced plateaus discussed in the SM. 

\begin{figure}[t]
\centering \includegraphics[width=0.9\linewidth]{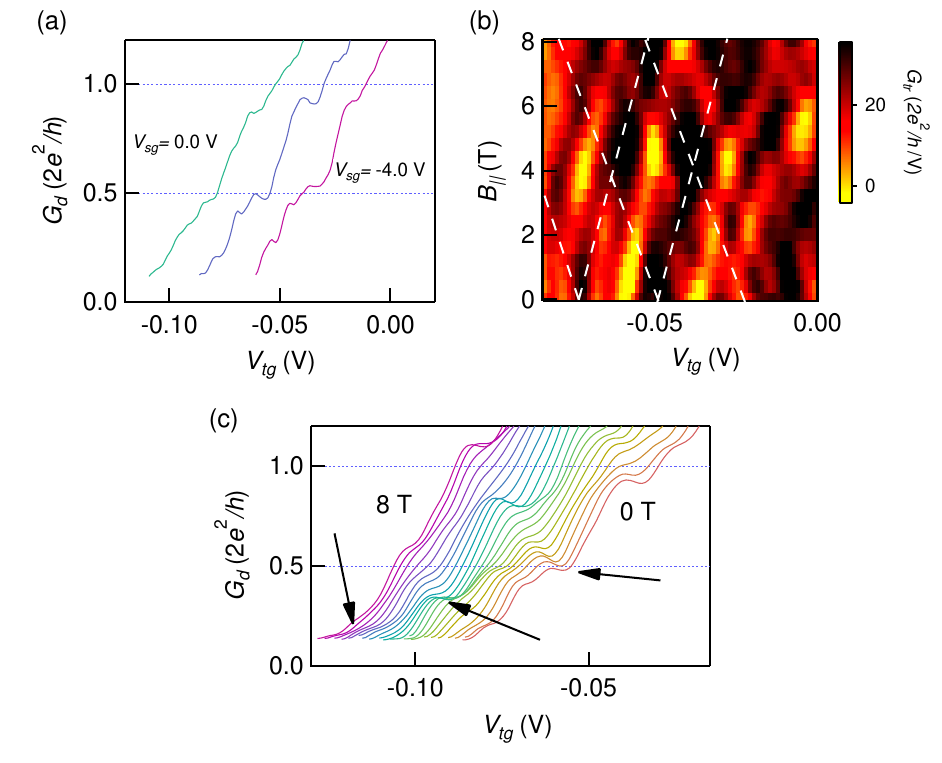}
\caption{(a) $G_d$ vs. $V_{tg}$ of the 1.4 ${\rm \mu}$m-long wire device with different $V_{sg}$. Plateaus at 0.5 and 1.0$\times 2e^2/h$ are observed. (b)Trans-conductance as functions of $B_{\parallel}$ and $V_{tg}$. Diamond structure originating from the Zeeman splitting is found as similar to the case of the 0.8 ${\rm \mu}$m-long wire. (c) $G_d$ vs. $V_{tg}$ obtained at $0$~T$<B_{\parallel}<8$~T. The horizontal axis shows the result at $B_{\parallel}=0$ T and the other results are incrementally shifted by -0.002 V for clarity. The plateau conductance at finite magnetic field is lower than the $2e^2/h$, as highlighted by arrows indicating the plateau at 8 T, 4 T and 0 T.}
\label{fig4} 
\end{figure}
Finally we measure a longer wire device (1.4 ${\rm \mu}$m length) to check the reproducibility of the HQP data. Figure~\ref{fig4}(a) indicates the $G_d$ vs. $V_{tg}$ obtained at $T =1.5$ K in the case of $V_{sg}=0,$~-2.0 and -4.0 V, respectively. Zero field HQP appears at $G_d =0.5\times2e^{2}/h$ in addition to the quantized plateau at $G_d =2e^{2}/h$. Figure~\ref{fig4}(b) shows $G_{tr}$ vs. $V_{tg}$ with $V_{sg}=-2.0$ V measured as functions of $B_{\parallel}$ and $V_{tg}$. From this dependency, we can estimate the Lande g factor of 5.6, which is consistent with the estimated value in the 0.8 ${\rm \mu}$m-long wire devices. Figure~\ref{fig4}(c) shows the $G_d$ vs. $V_{tg}$ at 0 T$\leq B_{\parallel}\leq8$ T. In Fig.~\ref{fig4}(c) $G_d$ of the plateaus at finite magnetic fields as indicated by two arrows for 4 and 8 T is smaller than that at 0 T. The main features of HQPs observed in the 0.8 ${\rm \mu}$m-length wire device are all reproduced in the longer wire device, and therefore we conclude that this anomalous conductance reduction observed at high fields is not due to impurities or roughness in the wires but an intrinsic phenomenon. 

We first discuss the observed features in the context of predictions from SSP and SG. Both proposed mechanisms are based on strong SOI predicting to generate HQPs previously observed in InAs or InGaAs quantum wires. Firstly, SSP is not suitable to explain our results, especially the Zeeman splitting or the zero field HSQ features. The observed Zeeman splitting is strong evidence that HQPs consist of spin-degenerated subbands. Additionally, the proposed spin polarization mechanism requires a lateral electric field induced by asymmetry in the side gating, whereas we observe the zero field HQPs even at $V_{sg}=0$~V. SG also does not hold for our results. In this mechanism, the subbands are spin-degenerated but the transmittance of the wire depends on the spin direction perpendicular to the plane of the injected electrons. Therefore it is expected that the subband energy of HQP decreases when $B_{\perp}$ is applied in the same direction as the injected electron spin, while the energy increases when $B_{\perp}$ applied in the opposite direction. The $B_{\perp}$ dependence in the DC bias measurement should be asymmetry about $B_{\perp}$=0 T. However, we do not recognize any such asymmetry as shown in Fig. \ref{fig2}(c). 

For several reasons we conclude that HQPs and the anomalous reduction of the HQP conductance with magnetic field arise from e-e interaction. First, there is no significant difference between the $B_{\parallel}$ dependence and the $B_{\perp}$ dependence, in contrast to expectations for spin-related phenomena. Secondly, plateau conductance significantly below $e^2/h$ is difficult to explain with a single electron picture which would result in $e^2/h$ as the lowest plateau conductance in the ballistic transport regime. At least, we can conclude that the magnetic field dependences are not explained by SSP and SG, namely the mechanisms with strong SOI in the wire. Therefore, we suspect that the observed features are assigned to SILL, the mechanism induced by the e-e interaction. However SILL canft be assigned to the anomalous conductance reduction induced by the in-plane magnetic fields. In order to completely reveal the mechanism, further theoretical and experimental efforts are required. 

In summary, we experimentally studied the electron transport in InAs quantum wires and its in-plane magnetic field dependence. Using a high quality InAs quantum well with large g factor, we observed HQPs and the in-plane magnetic field dependence of the Zeeman splitting, indicating that HQPs are formed from spin-degenerated subbands. In addition, we discovered that the HQP plateau conductance decreases as the in-plane magnetic field increases. These results are inconsistent with predictions from the previously proposed mechanisms such as the spontaneous spin polarization and the Stern-Gerlach mechanism. Though not clarified we finally assume the e-e interaction as a possible mechanism to account for the observed HQP feature. 

\section*{Acknowledgment}
We greatly thank Yasuhiro Tokura, Makoto Kohda, and Peter Stano for fruitful discussions.
This work was partially supported by Grant-in-Aid for Young Scientific Research (A) (No. JP15H05407), Grant-in-Aid for Scientific Research (A) (No. JP16H02204), Grant-in-Aid for Scientific Research (S) (No. JP26220710), JSPS Research Fellowship for Young Scientists (No. JP14J10600), JSPS Program for Leading Graduate Schools (MERIT) from JSPS, Grant-in-Aid for Scientific Research on Innovative Area, "Nano Spin Conversion Science" (No.JP15H01012 and JP17H05177), Grant-in-Aid for Scientific Research on Innovative Area, "Topological Materials Science" (Grant No. JP16H00984) from MEXT, CREST(No. JPMJCR15N2), and the Murata Science Foundation.

\clearpage
\section*{Supplemental Material}
\subparagraph{Stack materials of the InAs quantum well wafer}

\begin{figure}[h]
\centering \includegraphics[width=0.4\linewidth]{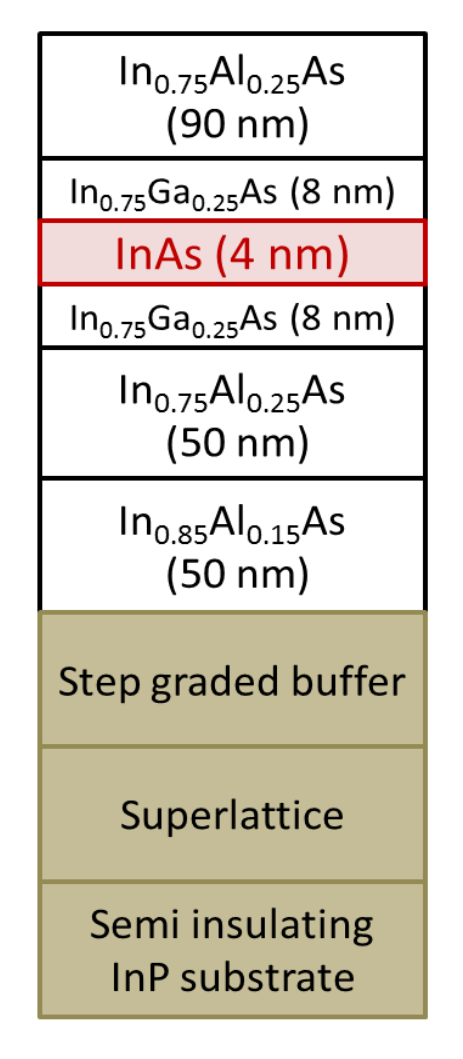} 
\caption{Schematic image of stack materials in the InAs quantum well wafer.}
\label{figs1} 
\end{figure}

Stack materials of our quantum well wafer are schematically shown in Fig.~\ref{figs1}. The 4 nm-thick InAs quantum well layer is embedded in the heterostructure.

\subparagraph{Temperature dependence of the zero field HQPs}

\begin{figure}[h]
\centering \includegraphics[width=0.7\linewidth]{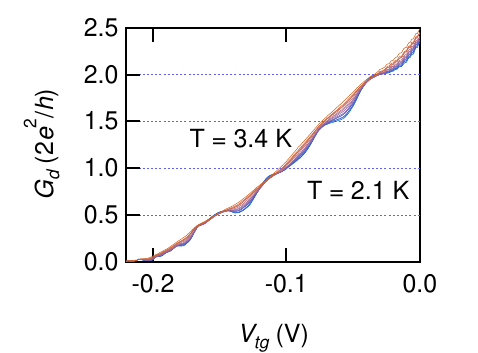} 
\caption{Temperature dependence of $G_d$ vs $V_{tg}$ for 2.1 K$<T<$3.4 K.}
\label{figs2} 
\end{figure}

Figure~\ref{figs2} shows the temperature dependence of $G_d$ measured in the range of  2.1 K$<T<$3.4 K. As $T$ increases, the plateau shape becomes less pronounced and finally vanishes. We expect that the conductance at the center of the plateau remains constant with temperature as long as the conductance plateau is well defined. However, Fig.~\ref{figs2} shows that the plateau conductance becomes significantly lower as the temperature decreases. We note that we also performed the measurements at $T=$50 mK, but for $T <$ 1 K non-periodic conductance fluctuations appear in $G_d$ vs $V_{tg}$ possibly due to quantum interference. This conductance fluctuation complicates the evaluation of the plateau conductance, and therefore we only discuss experimental results obtained at $T>1.5$ K here. The conductance decrease as temperature decreases is not expected in SSP and SG. According to SILL, the conductance decreases first to $e^2/h$ and then increase to $2e^2/h$ as temperature is decreased. Our temperature dependence is only measured above 1.5 K and so at least in this temperature range, the result follows the expected dependence of SILL.

\subparagraph{Bias voltage dependence of the zero field HQPs}

\begin{figure}[h]
\centering \includegraphics[width=0.7\linewidth]{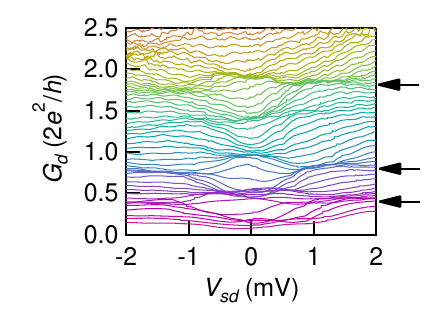} 
\caption{$G_d$ vs. bias voltage, $V_{sd}$ at different $V_{tg}$ between 0.185 V $< V_{tg} < 0.02$ V. The dense regions indicate the plateau structures. In addition to HQPs at $V_{sd}=0$ V, there are additional plateau structures at finite $V_{sd}$ indicated by the arrows.}
\label{figs3} 
\end{figure}

\begin{figure}[h]
\centering \includegraphics[width=0.7\linewidth]{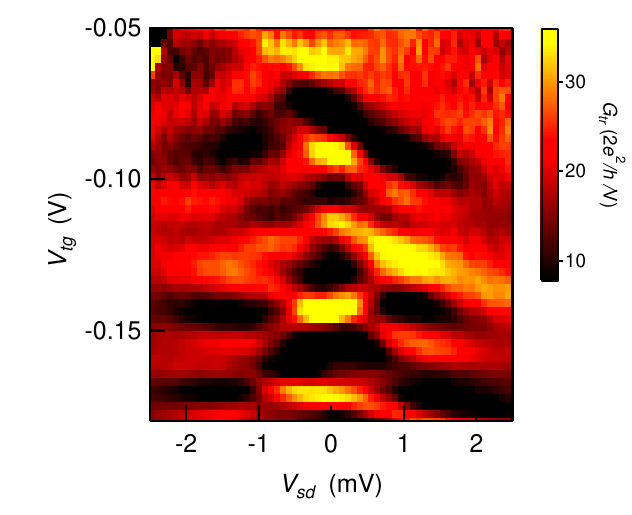} 
\caption{Trans-conductance, $G_{tr}$ as functions of $V_{sd}$ and $V_{tg}$. Diamond dependence from the subband structure is seen.}
\label{figs4} 
\end{figure}

We now focus on the results obtained at $T=$1.5 K. $G_d$ vs. bias voltage, $V_{sd}$, for 0.185 V $< V_{tg} < 0.02$ V obtained in the 1st cooldown is shown in Fig.~\ref{figs3}. The regions in which the curves are dense signify the conductance plateaus. In addition to zero field HQPs appearing at $V_{sd}=0$ V, the conductance plateaus are seen at finite $V_{sd}$. These plateaus observed around $0.35,~0.75~{\rm and}~1.75\times2e^{2}/h$ are indicated by arrows in Fig.~\ref{figs3}. They are assigned to a mechanism established in studies of GaAs quantum wires at finite bias voltages~\cite{pateljpcm1990, patelprb1991, thomasapl1995}: a subband whose bottom is in the bias window has half the contribution of other subbands whose bottoms are below the bias window. These induced plateaus can be observed at $0.25,~0.75,~1.25~{\rm and}~1.75\times2e^{2}/h$. Note they appear only in clean quantum wires because scattering events become more likely when more unoccupied subbands are energetically available at larger $V_{sd}$~\cite{rosslernjp2011}. Therefore, we can ignore the defects or impurity effect as the origin of HQPs. 
From the bias measurement, we calculate the energy difference between the subbands. Fig.~\ref{figs4} shows $G_{tr}$ as functions of $V_{sd}$ and $V_{tg}$. The dark regions correspond to the plateau structures. From the diamond-shapes corresponding to the plateau-transition regions, the inter-subband energy of 0.70 meV can be obtained. In addition, $V_{tg}$ length of $0.5\times2e^{2}/h$ is 0.029 V. These two values should be equal, so we can now decide a coefficient to convert $V_{tg}$ length, $\Delta V_{tg}$ to the energy in a unit of meV, $24\times \Delta V_{tg}$.
We note that the sizes of the three measured HQP diamonds (i.e. the subband energy differences) are nearly equal to 0.7 meV. Additionally we cannot identify zero bias conductance peaks at $0.5~{\rm or}~1.5\times2e^{2}/h$ plateaus in Fig.~\ref{figs3}. These features are qualitatively different from those reported for the 0.7 anomaly which shows sharp zero bias conductance peaks at the anomaly. 

\bibliographystyle{apsrev4-1}

\begin{thebibliography}{40}%
\makeatletter
\providecommand \@ifxundefined [1]{%
 \@ifx{#1\undefined}
}%
\providecommand \@ifnum [1]{%
 \ifnum #1\expandafter \@firstoftwo
 \else \expandafter \@secondoftwo
 \fi
}%
\providecommand \@ifx [1]{%
 \ifx #1\expandafter \@firstoftwo
 \else \expandafter \@secondoftwo
 \fi
}%
\providecommand \natexlab [1]{#1}%
\providecommand \enquote  [1]{``#1''}%
\providecommand \bibnamefont  [1]{#1}%
\providecommand \bibfnamefont [1]{#1}%
\providecommand \citenamefont [1]{#1}%
\providecommand \href@noop [0]{\@secondoftwo}%
\providecommand \href [0]{\begingroup \@sanitize@url \@href}%
\providecommand \@href[1]{\@@startlink{#1}\@@href}%
\providecommand \@@href[1]{\endgroup#1\@@endlink}%
\providecommand \@sanitize@url [0]{\catcode `\\12\catcode `\$12\catcode
  `\&12\catcode `\#12\catcode `\^12\catcode `\_12\catcode `\%12\relax}%
\providecommand \@@startlink[1]{}%
\providecommand \@@endlink[0]{}%
\providecommand \url  [0]{\begingroup\@sanitize@url \@url }%
\providecommand \@url [1]{\endgroup\@href {#1}{\urlprefix }}%
\providecommand \urlprefix  [0]{URL }%
\providecommand \Eprint [0]{\href }%
\providecommand \doibase [0]{http://dx.doi.org/}%
\providecommand \selectlanguage [0]{\@gobble}%
\providecommand \bibinfo  [0]{\@secondoftwo}%
\providecommand \bibfield  [0]{\@secondoftwo}%
\providecommand \translation [1]{[#1]}%
\providecommand \BibitemOpen [0]{}%
\providecommand \bibitemStop [0]{}%
\providecommand \bibitemNoStop [0]{.\EOS\space}%
\providecommand \EOS [0]{\spacefactor3000\relax}%
\providecommand \BibitemShut  [1]{\csname bibitem#1\endcsname}%
\let\auto@bib@innerbib\@empty
\bibitem [{\citenamefont {van Wees}\ \emph {et~al.}(1988)\citenamefont {van
  Wees}, \citenamefont {van Houten}, \citenamefont {Beenakker}, \citenamefont
  {Williamson}, \citenamefont {Kouwenhoven}, \citenamefont {van~der Marel},\
  and\ \citenamefont {Foxon}}]{weesprl1988}%
  \BibitemOpen
  \bibfield  {author} {\bibinfo {author} {\bibfnamefont {B.~J.}\ \bibnamefont
  {van Wees}}, \bibinfo {author} {\bibfnamefont {H.}~\bibnamefont {van
  Houten}}, \bibinfo {author} {\bibfnamefont {C.~W.~J.}\ \bibnamefont
  {Beenakker}}, \bibinfo {author} {\bibfnamefont {J.~G.}\ \bibnamefont
  {Williamson}}, \bibinfo {author} {\bibfnamefont {L.~P.}\ \bibnamefont
  {Kouwenhoven}}, \bibinfo {author} {\bibfnamefont {D.}~\bibnamefont {van~der
  Marel}}, \ and\ \bibinfo {author} {\bibfnamefont {C.~T.}\ \bibnamefont
  {Foxon}},\ }\href {\doibase 10.1103/PhysRevLett.60.848} {\bibfield  {journal}
  {\bibinfo  {journal} {Phys. Rev. Lett.}\ }\textbf {\bibinfo {volume} {60}},\
  \bibinfo {pages} {848} (\bibinfo {year} {1988})}\BibitemShut {NoStop}%
\bibitem [{\citenamefont {Tarucha}\ \emph {et~al.}(1995)\citenamefont
  {Tarucha}, \citenamefont {Honda},\ and\ \citenamefont
  {Saku}}]{taruchassc1995}%
  \BibitemOpen
  \bibfield  {author} {\bibinfo {author} {\bibfnamefont {S.}~\bibnamefont
  {Tarucha}}, \bibinfo {author} {\bibfnamefont {T.}~\bibnamefont {Honda}}, \
  and\ \bibinfo {author} {\bibfnamefont {T.}~\bibnamefont {Saku}},\ }\href@noop
  {} {\bibfield  {journal} {\bibinfo  {journal} {Solid State Communications}\
  }\textbf {\bibinfo {volume} {94}},\ \bibinfo {pages} {413} (\bibinfo {year}
  {1995})}\BibitemShut {NoStop}%
\bibitem [{\citenamefont {Yacoby}\ \emph {et~al.}(1996)\citenamefont {Yacoby},
  \citenamefont {Stormer}, \citenamefont {Wingreen}, \citenamefont {Pfeiffer},
  \citenamefont {Baldwin},\ and\ \citenamefont {West}}]{yacobyprl1996}%
  \BibitemOpen
  \bibfield  {author} {\bibinfo {author} {\bibfnamefont {A.}~\bibnamefont
  {Yacoby}}, \bibinfo {author} {\bibfnamefont {H.~L.}\ \bibnamefont {Stormer}},
  \bibinfo {author} {\bibfnamefont {N.~S.}\ \bibnamefont {Wingreen}}, \bibinfo
  {author} {\bibfnamefont {L.~N.}\ \bibnamefont {Pfeiffer}}, \bibinfo {author}
  {\bibfnamefont {K.~W.}\ \bibnamefont {Baldwin}}, \ and\ \bibinfo {author}
  {\bibfnamefont {K.~W.}\ \bibnamefont {West}},\ }\href {\doibase
  10.1103/PhysRevLett.77.4612} {\bibfield  {journal} {\bibinfo  {journal}
  {Phys. Rev. Lett.}\ }\textbf {\bibinfo {volume} {77}},\ \bibinfo {pages}
  {4612} (\bibinfo {year} {1996})}\BibitemShut {NoStop}%
\bibitem [{\citenamefont {Levy}\ \emph {et~al.}(2006)\citenamefont {Levy},
  \citenamefont {Tsukernik}, \citenamefont {Karpovski}, \citenamefont
  {Palevski}, \citenamefont {Dwir}, \citenamefont {Pelucchi}, \citenamefont
  {Rudra}, \citenamefont {Kapon},\ and\ \citenamefont {Oreg}}]{levyprl2006}%
  \BibitemOpen
  \bibfield  {author} {\bibinfo {author} {\bibfnamefont {E.}~\bibnamefont
  {Levy}}, \bibinfo {author} {\bibfnamefont {A.}~\bibnamefont {Tsukernik}},
  \bibinfo {author} {\bibfnamefont {M.}~\bibnamefont {Karpovski}}, \bibinfo
  {author} {\bibfnamefont {A.}~\bibnamefont {Palevski}}, \bibinfo {author}
  {\bibfnamefont {B.}~\bibnamefont {Dwir}}, \bibinfo {author} {\bibfnamefont
  {E.}~\bibnamefont {Pelucchi}}, \bibinfo {author} {\bibfnamefont
  {A.}~\bibnamefont {Rudra}}, \bibinfo {author} {\bibfnamefont
  {E.}~\bibnamefont {Kapon}}, \ and\ \bibinfo {author} {\bibfnamefont
  {Y.}~\bibnamefont {Oreg}},\ }\href {\doibase 10.1103/PhysRevLett.97.196802}
  {\bibfield  {journal} {\bibinfo  {journal} {Phys. Rev. Lett.}\ }\textbf
  {\bibinfo {volume} {97}},\ \bibinfo {pages} {196802} (\bibinfo {year}
  {2006})}\BibitemShut {NoStop}%
\bibitem [{\citenamefont {Tomonaga}(1950)}]{tomonagaptp1950}%
  \BibitemOpen
  \bibfield  {author} {\bibinfo {author} {\bibfnamefont {S.}~\bibnamefont
  {Tomonaga}},\ }\href@noop {} {\bibfield  {journal} {\bibinfo  {journal}
  {Prog. Theor. Phys.}\ }\textbf {\bibinfo {volume} {5}},\ \bibinfo {pages}
  {544} (\bibinfo {year} {1950})}\BibitemShut {NoStop}%
\bibitem [{\citenamefont {Kane}\ and\ \citenamefont
  {Fisher}(1992)}]{kaneprb1992}%
  \BibitemOpen
  \bibfield  {author} {\bibinfo {author} {\bibfnamefont {C.~L.}\ \bibnamefont
  {Kane}}\ and\ \bibinfo {author} {\bibfnamefont {M.~P.~A.}\ \bibnamefont
  {Fisher}},\ }\href {\doibase 10.1103/PhysRevB.46.15233} {\bibfield  {journal}
  {\bibinfo  {journal} {Phys. Rev. B}\ }\textbf {\bibinfo {volume} {46}},\
  \bibinfo {pages} {15233} (\bibinfo {year} {1992})}\BibitemShut {NoStop}%
\bibitem [{\citenamefont {Ogata}\ and\ \citenamefont
  {Fukuyama}(1994)}]{ogataprl1994}%
  \BibitemOpen
  \bibfield  {author} {\bibinfo {author} {\bibfnamefont {M.}~\bibnamefont
  {Ogata}}\ and\ \bibinfo {author} {\bibfnamefont {H.}~\bibnamefont
  {Fukuyama}},\ }\href {\doibase 10.1103/PhysRevLett.73.468} {\bibfield
  {journal} {\bibinfo  {journal} {Phys. Rev. Lett.}\ }\textbf {\bibinfo
  {volume} {73}},\ \bibinfo {pages} {468} (\bibinfo {year} {1994})}\BibitemShut
  {NoStop}%
\bibitem [{\citenamefont {Maslov}(1995)}]{maslovprb1995}%
  \BibitemOpen
  \bibfield  {author} {\bibinfo {author} {\bibfnamefont {D.~L.}\ \bibnamefont
  {Maslov}},\ }\href {\doibase 10.1103/PhysRevB.52.R14368} {\bibfield
  {journal} {\bibinfo  {journal} {Phys. Rev. B}\ }\textbf {\bibinfo {volume}
  {52}},\ \bibinfo {pages} {R14368} (\bibinfo {year} {1995})}\BibitemShut
  {NoStop}%
\bibitem [{\citenamefont {Thomas}\ \emph {et~al.}(1996)\citenamefont {Thomas},
  \citenamefont {Nicholls}, \citenamefont {Simmons}, \citenamefont {Pepper},
  \citenamefont {Mace},\ and\ \citenamefont {Ritchie}}]{thomasprl1996}%
  \BibitemOpen
  \bibfield  {author} {\bibinfo {author} {\bibfnamefont {K.~J.}\ \bibnamefont
  {Thomas}}, \bibinfo {author} {\bibfnamefont {J.~T.}\ \bibnamefont
  {Nicholls}}, \bibinfo {author} {\bibfnamefont {M.~Y.}\ \bibnamefont
  {Simmons}}, \bibinfo {author} {\bibfnamefont {M.}~\bibnamefont {Pepper}},
  \bibinfo {author} {\bibfnamefont {D.~R.}\ \bibnamefont {Mace}}, \ and\
  \bibinfo {author} {\bibfnamefont {D.~A.}\ \bibnamefont {Ritchie}},\ }\href
  {\doibase 10.1103/PhysRevLett.77.135} {\bibfield  {journal} {\bibinfo
  {journal} {Phys. Rev. Lett.}\ }\textbf {\bibinfo {volume} {77}},\ \bibinfo
  {pages} {135} (\bibinfo {year} {1996})}\BibitemShut {NoStop}%
\bibitem [{\citenamefont {Thomas}\ \emph {et~al.}(1998)\citenamefont {Thomas},
  \citenamefont {Nicholls}, \citenamefont {Appleyard}, \citenamefont {Simmons},
  \citenamefont {Pepper}, \citenamefont {Mace}, \citenamefont {Tribe},\ and\
  \citenamefont {Ritchie}}]{thomasprb1998}%
  \BibitemOpen
  \bibfield  {author} {\bibinfo {author} {\bibfnamefont {K.~J.}\ \bibnamefont
  {Thomas}}, \bibinfo {author} {\bibfnamefont {J.~T.}\ \bibnamefont
  {Nicholls}}, \bibinfo {author} {\bibfnamefont {N.~J.}\ \bibnamefont
  {Appleyard}}, \bibinfo {author} {\bibfnamefont {M.~Y.}\ \bibnamefont
  {Simmons}}, \bibinfo {author} {\bibfnamefont {M.}~\bibnamefont {Pepper}},
  \bibinfo {author} {\bibfnamefont {D.~R.}\ \bibnamefont {Mace}}, \bibinfo
  {author} {\bibfnamefont {W.~R.}\ \bibnamefont {Tribe}}, \ and\ \bibinfo
  {author} {\bibfnamefont {D.~A.}\ \bibnamefont {Ritchie}},\ }\href {\doibase
  10.1103/PhysRevB.58.4846} {\bibfield  {journal} {\bibinfo  {journal} {Phys.
  Rev. B}\ }\textbf {\bibinfo {volume} {58}},\ \bibinfo {pages} {4846}
  (\bibinfo {year} {1998})}\BibitemShut {NoStop}%
\bibitem [{\citenamefont {Kristensen}\ \emph {et~al.}(2000)\citenamefont
  {Kristensen}, \citenamefont {Bruus}, \citenamefont {Hansen}, \citenamefont
  {Jensen}, \citenamefont {Lindelof}, \citenamefont {Marckmann}, \citenamefont
  {Nyg\aa{}rd}, \citenamefont {S\o{}rensen}, \citenamefont {Beuscher},
  \citenamefont {Forchel},\ and\ \citenamefont {Michel}}]{kristensenprb2000}%
  \BibitemOpen
  \bibfield  {author} {\bibinfo {author} {\bibfnamefont {A.}~\bibnamefont
  {Kristensen}}, \bibinfo {author} {\bibfnamefont {H.}~\bibnamefont {Bruus}},
  \bibinfo {author} {\bibfnamefont {A.~E.}\ \bibnamefont {Hansen}}, \bibinfo
  {author} {\bibfnamefont {J.~B.}\ \bibnamefont {Jensen}}, \bibinfo {author}
  {\bibfnamefont {P.~E.}\ \bibnamefont {Lindelof}}, \bibinfo {author}
  {\bibfnamefont {C.~J.}\ \bibnamefont {Marckmann}}, \bibinfo {author}
  {\bibfnamefont {J.}~\bibnamefont {Nyg\aa{}rd}}, \bibinfo {author}
  {\bibfnamefont {C.~B.}\ \bibnamefont {S\o{}rensen}}, \bibinfo {author}
  {\bibfnamefont {F.}~\bibnamefont {Beuscher}}, \bibinfo {author}
  {\bibfnamefont {A.}~\bibnamefont {Forchel}}, \ and\ \bibinfo {author}
  {\bibfnamefont {M.}~\bibnamefont {Michel}},\ }\href {\doibase
  10.1103/PhysRevB.62.10950} {\bibfield  {journal} {\bibinfo  {journal} {Phys.
  Rev. B}\ }\textbf {\bibinfo {volume} {62}},\ \bibinfo {pages} {10950}
  (\bibinfo {year} {2000})}\BibitemShut {NoStop}%
\bibitem [{\citenamefont {Reilly}\ \emph {et~al.}(2001)\citenamefont {Reilly},
  \citenamefont {Facer}, \citenamefont {Dzurak}, \citenamefont {Kane},
  \citenamefont {Clark}, \citenamefont {Stiles}, \citenamefont {Clark},
  \citenamefont {Hamilton}, \citenamefont {O'Brien}, \citenamefont {Lumpkin},
  \citenamefont {Pfeiffer},\ and\ \citenamefont {West}}]{reillyprb2001}%
  \BibitemOpen
  \bibfield  {author} {\bibinfo {author} {\bibfnamefont {D.~J.}\ \bibnamefont
  {Reilly}}, \bibinfo {author} {\bibfnamefont {G.~R.}\ \bibnamefont {Facer}},
  \bibinfo {author} {\bibfnamefont {A.~S.}\ \bibnamefont {Dzurak}}, \bibinfo
  {author} {\bibfnamefont {B.~E.}\ \bibnamefont {Kane}}, \bibinfo {author}
  {\bibfnamefont {R.~G.}\ \bibnamefont {Clark}}, \bibinfo {author}
  {\bibfnamefont {P.~J.}\ \bibnamefont {Stiles}}, \bibinfo {author}
  {\bibfnamefont {R.~G.}\ \bibnamefont {Clark}}, \bibinfo {author}
  {\bibfnamefont {A.~R.}\ \bibnamefont {Hamilton}}, \bibinfo {author}
  {\bibfnamefont {J.~L.}\ \bibnamefont {O'Brien}}, \bibinfo {author}
  {\bibfnamefont {N.~E.}\ \bibnamefont {Lumpkin}}, \bibinfo {author}
  {\bibfnamefont {L.~N.}\ \bibnamefont {Pfeiffer}}, \ and\ \bibinfo {author}
  {\bibfnamefont {K.~W.}\ \bibnamefont {West}},\ }\href {\doibase
  10.1103/PhysRevB.63.121311} {\bibfield  {journal} {\bibinfo  {journal} {Phys.
  Rev. B}\ }\textbf {\bibinfo {volume} {63}},\ \bibinfo {pages} {121311}
  (\bibinfo {year} {2001})}\BibitemShut {NoStop}%
\bibitem [{\citenamefont {Cronenwett}\ \emph {et~al.}(2002)\citenamefont
  {Cronenwett}, \citenamefont {Lynch}, \citenamefont {Goldhaber-Gordon},
  \citenamefont {Kouwenhoven}, \citenamefont {Marcus}, \citenamefont {Hirose},
  \citenamefont {Wingreen},\ and\ \citenamefont {Umansky}}]{cronenwettprl2002}%
  \BibitemOpen
  \bibfield  {author} {\bibinfo {author} {\bibfnamefont {S.~M.}\ \bibnamefont
  {Cronenwett}}, \bibinfo {author} {\bibfnamefont {H.~J.}\ \bibnamefont
  {Lynch}}, \bibinfo {author} {\bibfnamefont {D.}~\bibnamefont
  {Goldhaber-Gordon}}, \bibinfo {author} {\bibfnamefont {L.~P.}\ \bibnamefont
  {Kouwenhoven}}, \bibinfo {author} {\bibfnamefont {C.~M.}\ \bibnamefont
  {Marcus}}, \bibinfo {author} {\bibfnamefont {K.}~\bibnamefont {Hirose}},
  \bibinfo {author} {\bibfnamefont {N.~S.}\ \bibnamefont {Wingreen}}, \ and\
  \bibinfo {author} {\bibfnamefont {V.}~\bibnamefont {Umansky}},\ }\href
  {\doibase 10.1103/PhysRevLett.88.226805} {\bibfield  {journal} {\bibinfo
  {journal} {Phys. Rev. Lett.}\ }\textbf {\bibinfo {volume} {88}},\ \bibinfo
  {pages} {226805} (\bibinfo {year} {2002})}\BibitemShut {NoStop}%
\bibitem [{\citenamefont {Rokhinson}\ \emph {et~al.}(2006)\citenamefont
  {Rokhinson}, \citenamefont {Pfeiffer},\ and\ \citenamefont
  {West}}]{rokhinsonprl2006}%
  \BibitemOpen
  \bibfield  {author} {\bibinfo {author} {\bibfnamefont {L.~P.}\ \bibnamefont
  {Rokhinson}}, \bibinfo {author} {\bibfnamefont {L.~N.}\ \bibnamefont
  {Pfeiffer}}, \ and\ \bibinfo {author} {\bibfnamefont {K.~W.}\ \bibnamefont
  {West}},\ }\href {\doibase 10.1103/PhysRevLett.96.156602} {\bibfield
  {journal} {\bibinfo  {journal} {Phys. Rev. Lett.}\ }\textbf {\bibinfo
  {volume} {96}},\ \bibinfo {pages} {156602} (\bibinfo {year}
  {2006})}\BibitemShut {NoStop}%
\bibitem [{\citenamefont {DiCarlo}\ \emph {et~al.}(2006)\citenamefont
  {DiCarlo}, \citenamefont {Zhang}, \citenamefont {McClure}, \citenamefont
  {Reilly}, \citenamefont {Marcus}, \citenamefont {Pfeiffer},\ and\
  \citenamefont {West}}]{dicarloprl2006}%
  \BibitemOpen
  \bibfield  {author} {\bibinfo {author} {\bibfnamefont {L.}~\bibnamefont
  {DiCarlo}}, \bibinfo {author} {\bibfnamefont {Y.}~\bibnamefont {Zhang}},
  \bibinfo {author} {\bibfnamefont {D.~T.}\ \bibnamefont {McClure}}, \bibinfo
  {author} {\bibfnamefont {D.~J.}\ \bibnamefont {Reilly}}, \bibinfo {author}
  {\bibfnamefont {C.~M.}\ \bibnamefont {Marcus}}, \bibinfo {author}
  {\bibfnamefont {L.~N.}\ \bibnamefont {Pfeiffer}}, \ and\ \bibinfo {author}
  {\bibfnamefont {K.~W.}\ \bibnamefont {West}},\ }\href {\doibase
  10.1103/PhysRevLett.97.036810} {\bibfield  {journal} {\bibinfo  {journal}
  {Phys. Rev. Lett.}\ }\textbf {\bibinfo {volume} {97}},\ \bibinfo {pages}
  {036810} (\bibinfo {year} {2006})}\BibitemShut {NoStop}%
\bibitem [{\citenamefont {Bauer}\ \emph {et~al.}(2013)\citenamefont {Bauer},
  \citenamefont {Heyder}, \citenamefont {Schubert}, \citenamefont {Borowsky},
  \citenamefont {Taubert}, \citenamefont {Bruognolo}, \citenamefont {Schuh},
  \citenamefont {Wegscheider}, \citenamefont {Delft},\ and\ \citenamefont
  {Ludwig}}]{bauernature2013}%
  \BibitemOpen
  \bibfield  {author} {\bibinfo {author} {\bibfnamefont {F.}~\bibnamefont
  {Bauer}}, \bibinfo {author} {\bibfnamefont {J.}~\bibnamefont {Heyder}},
  \bibinfo {author} {\bibfnamefont {E.}~\bibnamefont {Schubert}}, \bibinfo
  {author} {\bibfnamefont {D.}~\bibnamefont {Borowsky}}, \bibinfo {author}
  {\bibfnamefont {D.}~\bibnamefont {Taubert}}, \bibinfo {author} {\bibfnamefont
  {B.}~\bibnamefont {Bruognolo}}, \bibinfo {author} {\bibfnamefont
  {D.}~\bibnamefont {Schuh}}, \bibinfo {author} {\bibfnamefont
  {W.}~\bibnamefont {Wegscheider}}, \bibinfo {author} {\bibfnamefont {J.~v.}\
  \bibnamefont {Delft}}, \ and\ \bibinfo {author} {\bibfnamefont
  {S.}~\bibnamefont {Ludwig}},\ }\href@noop {} {\bibfield  {journal} {\bibinfo
  {journal} {Nature}\ }\textbf {\bibinfo {volume} {501}},\ \bibinfo {pages}
  {73} (\bibinfo {year} {2013})}\BibitemShut {NoStop}%
\bibitem [{\citenamefont {Iqbal}\ \emph {et~al.}(2013)\citenamefont {Iqbal},
  \citenamefont {Levy}, \citenamefont {Koop}, \citenamefont {Dekker},
  \citenamefont {de~Jong}, \citenamefont {van~der Velde}, \citenamefont
  {Reuter}, \citenamefont {Wieck}, \citenamefont {Aguado}, \citenamefont
  {Meir},\ and\ \citenamefont {van~der Wal}}]{iqbalnature2013}%
  \BibitemOpen
  \bibfield  {author} {\bibinfo {author} {\bibfnamefont {M.~J.}\ \bibnamefont
  {Iqbal}}, \bibinfo {author} {\bibfnamefont {R.}~\bibnamefont {Levy}},
  \bibinfo {author} {\bibfnamefont {E.~J.}\ \bibnamefont {Koop}}, \bibinfo
  {author} {\bibfnamefont {J.~B.}\ \bibnamefont {Dekker}}, \bibinfo {author}
  {\bibfnamefont {J.~P.}\ \bibnamefont {de~Jong}}, \bibinfo {author}
  {\bibfnamefont {J.~H.~M.}\ \bibnamefont {van~der Velde}}, \bibinfo {author}
  {\bibfnamefont {D.}~\bibnamefont {Reuter}}, \bibinfo {author} {\bibfnamefont
  {A.~D.}\ \bibnamefont {Wieck}}, \bibinfo {author} {\bibfnamefont
  {R.}~\bibnamefont {Aguado}}, \bibinfo {author} {\bibfnamefont
  {Y.}~\bibnamefont {Meir}}, \ and\ \bibinfo {author} {\bibfnamefont {C.~H.}\
  \bibnamefont {van~der Wal}},\ }\href@noop {} {\bibfield  {journal} {\bibinfo
  {journal} {Nature}\ }\textbf {\bibinfo {volume} {501}},\ \bibinfo {pages}
  {79?83} (\bibinfo {year} {2013})}\BibitemShut {NoStop}%
\bibitem [{\citenamefont {Crook}\ \emph {et~al.}(2006)\citenamefont {Crook},
  \citenamefont {Prance}, \citenamefont {Thomas}, \citenamefont {Chorley},
  \citenamefont {Farrer}, \citenamefont {Ritchie}, \citenamefont {Pepper},\
  and\ \citenamefont {Smith}}]{crookscience2006}%
  \BibitemOpen
  \bibfield  {author} {\bibinfo {author} {\bibfnamefont {R.}~\bibnamefont
  {Crook}}, \bibinfo {author} {\bibfnamefont {J.}~\bibnamefont {Prance}},
  \bibinfo {author} {\bibfnamefont {K.~J.}\ \bibnamefont {Thomas}}, \bibinfo
  {author} {\bibfnamefont {S.~J.}\ \bibnamefont {Chorley}}, \bibinfo {author}
  {\bibfnamefont {I.}~\bibnamefont {Farrer}}, \bibinfo {author} {\bibfnamefont
  {D.~A.}\ \bibnamefont {Ritchie}}, \bibinfo {author} {\bibfnamefont
  {M.}~\bibnamefont {Pepper}}, \ and\ \bibinfo {author} {\bibfnamefont {C.~G.}\
  \bibnamefont {Smith}},\ }\href@noop {} {\bibfield  {journal} {\bibinfo
  {journal} {Science}\ }\textbf {\bibinfo {volume} {312}},\ \bibinfo {pages}
  {1359} (\bibinfo {year} {2006})}\BibitemShut {NoStop}%
\bibitem [{\citenamefont {Hew}\ \emph {et~al.}(2008)\citenamefont {Hew},
  \citenamefont {Thomas}, \citenamefont {Pepper}, \citenamefont {Farrer},
  \citenamefont {Anderson}, \citenamefont {Jones},\ and\ \citenamefont
  {Ritchie}}]{hewpr;2008}%
  \BibitemOpen
  \bibfield  {author} {\bibinfo {author} {\bibfnamefont {W.~K.}\ \bibnamefont
  {Hew}}, \bibinfo {author} {\bibfnamefont {K.~J.}\ \bibnamefont {Thomas}},
  \bibinfo {author} {\bibfnamefont {M.}~\bibnamefont {Pepper}}, \bibinfo
  {author} {\bibfnamefont {I.}~\bibnamefont {Farrer}}, \bibinfo {author}
  {\bibfnamefont {D.}~\bibnamefont {Anderson}}, \bibinfo {author}
  {\bibfnamefont {G.~A.~C.}\ \bibnamefont {Jones}}, \ and\ \bibinfo {author}
  {\bibfnamefont {D.~A.}\ \bibnamefont {Ritchie}},\ }\href {\doibase
  10.1103/PhysRevLett.101.036801} {\bibfield  {journal} {\bibinfo  {journal}
  {Phys. Rev. Lett.}\ }\textbf {\bibinfo {volume} {101}},\ \bibinfo {pages}
  {036801} (\bibinfo {year} {2008})}\BibitemShut {NoStop}%
\bibitem [{\citenamefont {Biercuk}\ \emph {et~al.}(2005)\citenamefont
  {Biercuk}, \citenamefont {Mason}, \citenamefont {Martin}, \citenamefont
  {Yacoby},\ and\ \citenamefont {Marcus}}]{biercukprl2005}%
  \BibitemOpen
  \bibfield  {author} {\bibinfo {author} {\bibfnamefont {M.~J.}\ \bibnamefont
  {Biercuk}}, \bibinfo {author} {\bibfnamefont {N.}~\bibnamefont {Mason}},
  \bibinfo {author} {\bibfnamefont {J.}~\bibnamefont {Martin}}, \bibinfo
  {author} {\bibfnamefont {A.}~\bibnamefont {Yacoby}}, \ and\ \bibinfo {author}
  {\bibfnamefont {C.~M.}\ \bibnamefont {Marcus}},\ }\href {\doibase
  10.1103/PhysRevLett.94.026801} {\bibfield  {journal} {\bibinfo  {journal}
  {Phys. Rev. Lett.}\ }\textbf {\bibinfo {volume} {94}},\ \bibinfo {pages}
  {026801} (\bibinfo {year} {2005})}\BibitemShut {NoStop}%
\bibitem [{\citenamefont {Debray}\ \emph {et~al.}(2009)\citenamefont {Debray},
  \citenamefont {Rahman}, \citenamefont {Wan}, \citenamefont {Newrock},
  \citenamefont {Cahay}, \citenamefont {Ngo}, \citenamefont {Ulloa},
  \citenamefont {Herbert}, \citenamefont {Muhammad},\ and\ \citenamefont
  {Johnson}}]{debraynatnano2009}%
  \BibitemOpen
  \bibfield  {author} {\bibinfo {author} {\bibfnamefont {P.}~\bibnamefont
  {Debray}}, \bibinfo {author} {\bibfnamefont {S.~M.~S.}\ \bibnamefont
  {Rahman}}, \bibinfo {author} {\bibfnamefont {J.}~\bibnamefont {Wan}},
  \bibinfo {author} {\bibfnamefont {R.~S.}\ \bibnamefont {Newrock}}, \bibinfo
  {author} {\bibfnamefont {M.}~\bibnamefont {Cahay}}, \bibinfo {author}
  {\bibfnamefont {A.~T.}\ \bibnamefont {Ngo}}, \bibinfo {author} {\bibfnamefont
  {S.~E.}\ \bibnamefont {Ulloa}}, \bibinfo {author} {\bibfnamefont {S.~T.}\
  \bibnamefont {Herbert}}, \bibinfo {author} {\bibfnamefont {M.}~\bibnamefont
  {Muhammad}}, \ and\ \bibinfo {author} {\bibfnamefont {M.}~\bibnamefont
  {Johnson}},\ }\href@noop {} {\bibfield  {journal} {\bibinfo  {journal}
  {Nature Nanotechnology}\ }\textbf {\bibinfo {volume} {4}},\ \bibinfo {pages}
  {759} (\bibinfo {year} {2009})}\BibitemShut {NoStop}%
\bibitem [{\citenamefont {Wan}\ \emph {et~al.}(2009)\citenamefont {Wan},
  \citenamefont {Cahay}, \citenamefont {Debray},\ and\ \citenamefont
  {Newrock}}]{wanprb2009}%
  \BibitemOpen
  \bibfield  {author} {\bibinfo {author} {\bibfnamefont {J.}~\bibnamefont
  {Wan}}, \bibinfo {author} {\bibfnamefont {M.}~\bibnamefont {Cahay}}, \bibinfo
  {author} {\bibfnamefont {P.}~\bibnamefont {Debray}}, \ and\ \bibinfo {author}
  {\bibfnamefont {R.}~\bibnamefont {Newrock}},\ }\href {\doibase
  10.1103/PhysRevB.80.155440} {\bibfield  {journal} {\bibinfo  {journal} {Phys.
  Rev. B}\ }\textbf {\bibinfo {volume} {80}},\ \bibinfo {pages} {155440}
  (\bibinfo {year} {2009})}\BibitemShut {NoStop}%
\bibitem [{\citenamefont {Kohda}\ \emph {et~al.}(2012)\citenamefont {Kohda},
  \citenamefont {Nakamura}, \citenamefont {Nishihara}, \citenamefont
  {Kobayashi}, \citenamefont {Ono}, \citenamefont {Ohe}, \citenamefont
  {Tokura}, \citenamefont {Mineno},\ and\ \citenamefont
  {Nitta}}]{kohdanatcommun2012}%
  \BibitemOpen
  \bibfield  {author} {\bibinfo {author} {\bibfnamefont {M.}~\bibnamefont
  {Kohda}}, \bibinfo {author} {\bibfnamefont {S.}~\bibnamefont {Nakamura}},
  \bibinfo {author} {\bibfnamefont {Y.}~\bibnamefont {Nishihara}}, \bibinfo
  {author} {\bibfnamefont {K.}~\bibnamefont {Kobayashi}}, \bibinfo {author}
  {\bibfnamefont {T.}~\bibnamefont {Ono}}, \bibinfo {author} {\bibfnamefont
  {J.-i.}\ \bibnamefont {Ohe}}, \bibinfo {author} {\bibfnamefont
  {Y.}~\bibnamefont {Tokura}}, \bibinfo {author} {\bibfnamefont
  {T.}~\bibnamefont {Mineno}}, \ and\ \bibinfo {author} {\bibfnamefont
  {J.}~\bibnamefont {Nitta}},\ }\href@noop {} {\bibfield  {journal} {\bibinfo
  {journal} {Nature Communications}\ }\textbf {\bibinfo {volume} {3}},\
  \bibinfo {pages} {1082} (\bibinfo {year} {2012})}\BibitemShut {NoStop}%
\bibitem [{\citenamefont {Ngo}\ \emph {et~al.}(2010)\citenamefont {Ngo},
  \citenamefont {Debray},\ and\ \citenamefont {Ulloa}}]{ngoprb2010}%
  \BibitemOpen
  \bibfield  {author} {\bibinfo {author} {\bibfnamefont {A.~T.}\ \bibnamefont
  {Ngo}}, \bibinfo {author} {\bibfnamefont {P.}~\bibnamefont {Debray}}, \ and\
  \bibinfo {author} {\bibfnamefont {S.~E.}\ \bibnamefont {Ulloa}},\ }\href
  {\doibase 10.1103/PhysRevB.81.115328} {\bibfield  {journal} {\bibinfo
  {journal} {Phys. Rev. B}\ }\textbf {\bibinfo {volume} {81}},\ \bibinfo
  {pages} {115328} (\bibinfo {year} {2010})}\BibitemShut {NoStop}%
\bibitem [{\citenamefont {Matveev}(2004)}]{matveevprb2004}%
  \BibitemOpen
  \bibfield  {author} {\bibinfo {author} {\bibfnamefont {K.~A.}\ \bibnamefont
  {Matveev}},\ }\href {\doibase 10.1103/PhysRevB.70.245319} {\bibfield
  {journal} {\bibinfo  {journal} {Phys. Rev. B}\ }\textbf {\bibinfo {volume}
  {70}},\ \bibinfo {pages} {245319} (\bibinfo {year} {2004})}\BibitemShut
  {NoStop}%
\bibitem [{\citenamefont {Fiete}(2007)}]{fietermp2007}%
  \BibitemOpen
  \bibfield  {author} {\bibinfo {author} {\bibfnamefont {G.~A.}\ \bibnamefont
  {Fiete}},\ }\href {\doibase 10.1103/RevModPhys.79.801} {\bibfield  {journal}
  {\bibinfo  {journal} {Rev. Mod. Phys.}\ }\textbf {\bibinfo {volume} {79}},\
  \bibinfo {pages} {801} (\bibinfo {year} {2007})}\BibitemShut {NoStop}%
\bibitem [{\citenamefont {Fu}\ and\ \citenamefont {Kane}(2008)}]{fuprl2008}%
  \BibitemOpen
  \bibfield  {author} {\bibinfo {author} {\bibfnamefont {L.}~\bibnamefont
  {Fu}}\ and\ \bibinfo {author} {\bibfnamefont {C.~L.}\ \bibnamefont {Kane}},\
  }\href {\doibase 10.1103/PhysRevLett.100.096407} {\bibfield  {journal}
  {\bibinfo  {journal} {Phys. Rev. Lett.}\ }\textbf {\bibinfo {volume} {100}},\
  \bibinfo {pages} {096407} (\bibinfo {year} {2008})}\BibitemShut {NoStop}%
\bibitem [{\citenamefont {Hasan}\ and\ \citenamefont
  {Kane}(2010)}]{hasanrmp2010}%
  \BibitemOpen
  \bibfield  {author} {\bibinfo {author} {\bibfnamefont {M.~Z.}\ \bibnamefont
  {Hasan}}\ and\ \bibinfo {author} {\bibfnamefont {C.~L.}\ \bibnamefont
  {Kane}},\ }\href {\doibase 10.1103/RevModPhys.82.3045} {\bibfield  {journal}
  {\bibinfo  {journal} {Rev. Mod. Phys.}\ }\textbf {\bibinfo {volume} {82}},\
  \bibinfo {pages} {3045} (\bibinfo {year} {2010})}\BibitemShut {NoStop}%
\bibitem [{\citenamefont {Mourik}\ \emph {et~al.}(2012)\citenamefont {Mourik},
  \citenamefont {Zuo}, \citenamefont {Frolov}, \citenamefont {Plissard},
  \citenamefont {Bakkers},\ and\ \citenamefont
  {Kouwenhoven}}]{mourikscience2012}%
  \BibitemOpen
  \bibfield  {author} {\bibinfo {author} {\bibfnamefont {V.}~\bibnamefont
  {Mourik}}, \bibinfo {author} {\bibfnamefont {K.}~\bibnamefont {Zuo}},
  \bibinfo {author} {\bibfnamefont {S.~M.}\ \bibnamefont {Frolov}}, \bibinfo
  {author} {\bibfnamefont {S.~R.}\ \bibnamefont {Plissard}}, \bibinfo {author}
  {\bibfnamefont {E.~P. A.~M.}\ \bibnamefont {Bakkers}}, \ and\ \bibinfo
  {author} {\bibfnamefont {L.~P.}\ \bibnamefont {Kouwenhoven}},\ }\href
  {\doibase 10.1126/science.1222360} {\bibfield  {journal} {\bibinfo  {journal}
  {Science}\ }\textbf {\bibinfo {volume} {336}},\ \bibinfo {pages} {1003}
  (\bibinfo {year} {2012})}\BibitemShut {NoStop}%
\bibitem [{\citenamefont {Das}\ \emph {et~al.}(2012)\citenamefont {Das},
  \citenamefont {Ronen}, \citenamefont {Most}, \citenamefont {Oreg},
  \citenamefont {Heiblum},\ and\ \citenamefont {Shtrikman}}]{dasnatphys2012}%
  \BibitemOpen
  \bibfield  {author} {\bibinfo {author} {\bibfnamefont {A.}~\bibnamefont
  {Das}}, \bibinfo {author} {\bibfnamefont {Y.}~\bibnamefont {Ronen}}, \bibinfo
  {author} {\bibfnamefont {Y.}~\bibnamefont {Most}}, \bibinfo {author}
  {\bibfnamefont {Y.}~\bibnamefont {Oreg}}, \bibinfo {author} {\bibfnamefont
  {M.}~\bibnamefont {Heiblum}}, \ and\ \bibinfo {author} {\bibfnamefont
  {H.}~\bibnamefont {Shtrikman}},\ }\href@noop {} {\bibfield  {journal}
  {\bibinfo  {journal} {Nature Physics}\ }\textbf {\bibinfo {volume} {8}},\
  \bibinfo {pages} {887} (\bibinfo {year} {2012})}\BibitemShut {NoStop}%
\bibitem [{\citenamefont {Rokhinson}\ \emph {et~al.}(2012)\citenamefont
  {Rokhinson}, \citenamefont {Liu},\ and\ \citenamefont
  {Furdyna}}]{rokhinsonnatphys2012}%
  \BibitemOpen
  \bibfield  {author} {\bibinfo {author} {\bibfnamefont {L.~P.}\ \bibnamefont
  {Rokhinson}}, \bibinfo {author} {\bibfnamefont {X.}~\bibnamefont {Liu}}, \
  and\ \bibinfo {author} {\bibfnamefont {J.~K.}\ \bibnamefont {Furdyna}},\
  }\href@noop {} {\bibfield  {journal} {\bibinfo  {journal} {Nature Physics}\
  }\textbf {\bibinfo {volume} {8}},\ \bibinfo {pages} {795} (\bibinfo {year}
  {2012})}\BibitemShut {NoStop}%
\bibitem [{\citenamefont {Albrecht}\ \emph {et~al.}(2016)\citenamefont
  {Albrecht}, \citenamefont {Higginbotham}, \citenamefont {Madsen},
  \citenamefont {Kuemmeth}, \citenamefont {Jespersen}, \citenamefont {Nygard},
  \citenamefont {Krogstrup},\ and\ \citenamefont {Marcus}}]{albrechtnat2016}%
  \BibitemOpen
  \bibfield  {author} {\bibinfo {author} {\bibfnamefont {S.~M.}\ \bibnamefont
  {Albrecht}}, \bibinfo {author} {\bibfnamefont {A.~P.}\ \bibnamefont
  {Higginbotham}}, \bibinfo {author} {\bibfnamefont {M.}~\bibnamefont
  {Madsen}}, \bibinfo {author} {\bibfnamefont {F.}~\bibnamefont {Kuemmeth}},
  \bibinfo {author} {\bibfnamefont {T.~S.}\ \bibnamefont {Jespersen}}, \bibinfo
  {author} {\bibfnamefont {J.}~\bibnamefont {Nygard}}, \bibinfo {author}
  {\bibfnamefont {P.}~\bibnamefont {Krogstrup}}, \ and\ \bibinfo {author}
  {\bibfnamefont {C.~M.}\ \bibnamefont {Marcus}},\ }\href@noop {} {\bibfield
  {journal} {\bibinfo  {journal} {Nature}\ }\textbf {\bibinfo {volume} {531}},\
  \bibinfo {pages} {206} (\bibinfo {year} {2016})}\BibitemShut {NoStop}%
\bibitem [{\citenamefont {Shabani}\ \emph
  {et~al.}(2014{\natexlab{a}})\citenamefont {Shabani}, \citenamefont
  {Das~Sarma},\ and\ \citenamefont {Palmstr\o{}m}}]{shabaniprb2014}%
  \BibitemOpen
  \bibfield  {author} {\bibinfo {author} {\bibfnamefont {J.}~\bibnamefont
  {Shabani}}, \bibinfo {author} {\bibfnamefont {S.}~\bibnamefont {Das~Sarma}},
  \ and\ \bibinfo {author} {\bibfnamefont {C.~J.}\ \bibnamefont
  {Palmstr\o{}m}},\ }\href {\doibase 10.1103/PhysRevB.90.161303} {\bibfield
  {journal} {\bibinfo  {journal} {Phys. Rev. B}\ }\textbf {\bibinfo {volume}
  {90}},\ \bibinfo {pages} {161303} (\bibinfo {year}
  {2014}{\natexlab{a}})}\BibitemShut {NoStop}%
\bibitem [{\citenamefont {Shabani}\ \emph
  {et~al.}(2014{\natexlab{b}})\citenamefont {Shabani}, \citenamefont
  {McFadden}, \citenamefont {Shojaei},\ and\ \citenamefont
  {Palmstr\o{}m}}]{shabaniapl2014}%
  \BibitemOpen
  \bibfield  {author} {\bibinfo {author} {\bibfnamefont {J.}~\bibnamefont
  {Shabani}}, \bibinfo {author} {\bibfnamefont {A.~P.}\ \bibnamefont
  {McFadden}}, \bibinfo {author} {\bibfnamefont {B.}~\bibnamefont {Shojaei}}, \
  and\ \bibinfo {author} {\bibfnamefont {C.~J.}\ \bibnamefont {Palmstr\o{}m}},\
  }\href@noop {} {\bibfield  {journal} {\bibinfo  {journal} {Applied Physics
  Letters}\ }\textbf {\bibinfo {volume} {105}},\ \bibinfo {eid} {262105}
  (\bibinfo {year} {2014}{\natexlab{b}})}\BibitemShut {NoStop}%
\bibitem [{\citenamefont {Komijani}\ \emph {et~al.}(2013)\citenamefont
  {Komijani}, \citenamefont {Csontos}, \citenamefont {Shorubalko},
  \citenamefont {Z'[licke}, \citenamefont {Ihn}, \citenamefont {Ensslin},
  \citenamefont {Reuter},\ and\ \citenamefont {Wieck}}]{komijaniepl2013}%
  \BibitemOpen
  \bibfield  {author} {\bibinfo {author} {\bibfnamefont {Y.}~\bibnamefont
  {Komijani}}, \bibinfo {author} {\bibfnamefont {M.}~\bibnamefont {Csontos}},
  \bibinfo {author} {\bibfnamefont {I.}~\bibnamefont {Shorubalko}}, \bibinfo
  {author} {\bibfnamefont {U.}~\bibnamefont {Z'[licke}}, \bibinfo {author}
  {\bibfnamefont {T.}~\bibnamefont {Ihn}}, \bibinfo {author} {\bibfnamefont
  {K.}~\bibnamefont {Ensslin}}, \bibinfo {author} {\bibfnamefont
  {D.}~\bibnamefont {Reuter}}, \ and\ \bibinfo {author} {\bibfnamefont {A.~D.}\
  \bibnamefont {Wieck}},\ }\href
  {http://stacks.iop.org/0295-5075/102/i=3/a=37002} {\bibfield  {journal}
  {\bibinfo  {journal} {EPL (Europhysics Letters)}\ }\textbf {\bibinfo {volume}
  {102}},\ \bibinfo {pages} {37002} (\bibinfo {year} {2013})}\BibitemShut
  {NoStop}%
\bibitem [{\citenamefont {Nichele}\ \emph {et~al.}(2014)\citenamefont
  {Nichele}, \citenamefont {Chesi}, \citenamefont {Hennel}, \citenamefont
  {Wittmann}, \citenamefont {Gerl}, \citenamefont {Wegscheider}, \citenamefont
  {Loss}, \citenamefont {Ihn},\ and\ \citenamefont
  {Ensslin}}]{fabrizioprl2014}%
  \BibitemOpen
  \bibfield  {author} {\bibinfo {author} {\bibfnamefont {F.}~\bibnamefont
  {Nichele}}, \bibinfo {author} {\bibfnamefont {S.}~\bibnamefont {Chesi}},
  \bibinfo {author} {\bibfnamefont {S.}~\bibnamefont {Hennel}}, \bibinfo
  {author} {\bibfnamefont {A.}~\bibnamefont {Wittmann}}, \bibinfo {author}
  {\bibfnamefont {C.}~\bibnamefont {Gerl}}, \bibinfo {author} {\bibfnamefont
  {W.}~\bibnamefont {Wegscheider}}, \bibinfo {author} {\bibfnamefont
  {D.}~\bibnamefont {Loss}}, \bibinfo {author} {\bibfnamefont {T.}~\bibnamefont
  {Ihn}}, \ and\ \bibinfo {author} {\bibfnamefont {K.}~\bibnamefont
  {Ensslin}},\ }\href {\doibase 10.1103/PhysRevLett.113.046801} {\bibfield
  {journal} {\bibinfo  {journal} {Phys. Rev. Lett.}\ }\textbf {\bibinfo
  {volume} {113}},\ \bibinfo {pages} {046801} (\bibinfo {year}
  {2014})}\BibitemShut {NoStop}%
\bibitem [{\citenamefont {Patel}\ \emph {et~al.}(1990)\citenamefont {Patel},
  \citenamefont {Martin-Moreno}, \citenamefont {Pepper}, \citenamefont
  {Newbury}, \citenamefont {Frost}, \citenamefont {Ritchie}, \citenamefont
  {Jones}, \citenamefont {Janssen}, \citenamefont {JSingleton},\ and\
  \citenamefont {Perenboom}}]{pateljpcm1990}%
  \BibitemOpen
  \bibfield  {author} {\bibinfo {author} {\bibfnamefont {N.~K.}\ \bibnamefont
  {Patel}}, \bibinfo {author} {\bibfnamefont {L.}~\bibnamefont
  {Martin-Moreno}}, \bibinfo {author} {\bibfnamefont {M.}~\bibnamefont
  {Pepper}}, \bibinfo {author} {\bibfnamefont {R.}~\bibnamefont {Newbury}},
  \bibinfo {author} {\bibfnamefont {J.~E.~F.}\ \bibnamefont {Frost}}, \bibinfo
  {author} {\bibfnamefont {D.~A.}\ \bibnamefont {Ritchie}}, \bibinfo {author}
  {\bibfnamefont {G.~A.~C.}\ \bibnamefont {Jones}}, \bibinfo {author}
  {\bibfnamefont {J.~T. M.~B.}\ \bibnamefont {Janssen}}, \bibinfo {author}
  {\bibnamefont {JSingleton}}, \ and\ \bibinfo {author} {\bibfnamefont {J.~A.
  A.~J.}\ \bibnamefont {Perenboom}},\ }\href
  {http://stacks.iop.org/0953-8984/2/i=34/a=018} {\bibfield  {journal}
  {\bibinfo  {journal} {Journal of Physics: Condensed Matter}\ }\textbf
  {\bibinfo {volume} {2}},\ \bibinfo {pages} {7247} (\bibinfo {year}
  {1990})}\BibitemShut {NoStop}%
\bibitem [{\citenamefont {Patel}\ \emph {et~al.}(1991)\citenamefont {Patel},
  \citenamefont {Nicholls}, \citenamefont {Martn-Moreno}, \citenamefont
  {Pepper}, \citenamefont {Frost}, \citenamefont {Ritchie},\ and\ \citenamefont
  {Jones}}]{patelprb1991}%
  \BibitemOpen
  \bibfield  {author} {\bibinfo {author} {\bibfnamefont {N.~K.}\ \bibnamefont
  {Patel}}, \bibinfo {author} {\bibfnamefont {J.~T.}\ \bibnamefont {Nicholls}},
  \bibinfo {author} {\bibfnamefont {L.}~\bibnamefont {Martn-Moreno}}, \bibinfo
  {author} {\bibfnamefont {M.}~\bibnamefont {Pepper}}, \bibinfo {author}
  {\bibfnamefont {J.~E.~F.}\ \bibnamefont {Frost}}, \bibinfo {author}
  {\bibfnamefont {D.~A.}\ \bibnamefont {Ritchie}}, \ and\ \bibinfo {author}
  {\bibfnamefont {G.~A.~C.}\ \bibnamefont {Jones}},\ }\href {\doibase
  10.1103/PhysRevB.44.13549} {\bibfield  {journal} {\bibinfo  {journal} {Phys.
  Rev. B}\ }\textbf {\bibinfo {volume} {44}},\ \bibinfo {pages} {13549}
  (\bibinfo {year} {1991})}\BibitemShut {NoStop}%
\bibitem [{\citenamefont {Thomas}\ \emph {et~al.}(1995)\citenamefont {Thomas},
  \citenamefont {Simmons}, \citenamefont {Nicholls}, \citenamefont {Mace},
  \citenamefont {Pepper},\ and\ \citenamefont {Ritchie}}]{thomasapl1995}%
  \BibitemOpen
  \bibfield  {author} {\bibinfo {author} {\bibfnamefont {K.~J.}\ \bibnamefont
  {Thomas}}, \bibinfo {author} {\bibfnamefont {M.~Y.}\ \bibnamefont {Simmons}},
  \bibinfo {author} {\bibfnamefont {J.~T.}\ \bibnamefont {Nicholls}}, \bibinfo
  {author} {\bibfnamefont {D.~R.}\ \bibnamefont {Mace}}, \bibinfo {author}
  {\bibfnamefont {M.}~\bibnamefont {Pepper}}, \ and\ \bibinfo {author}
  {\bibfnamefont {D.~A.}\ \bibnamefont {Ritchie}},\ }\href {\doibase
  10.1063/1.115498} {\bibfield  {journal} {\bibinfo  {journal} {Applied Physics
  Letters}\ }\textbf {\bibinfo {volume} {67}},\ \bibinfo {pages} {109}
  (\bibinfo {year} {1995})},\ \Eprint
  {http://arxiv.org/abs/http://dx.doi.org/10.1063/1.115498}
  {http://dx.doi.org/10.1063/1.115498} \BibitemShut {NoStop}%
\bibitem [{\citenamefont {Rossler}\ \emph {et~al.}(2011)\citenamefont
  {Rossler}, \citenamefont {Baer}, \citenamefont {de~Wiljes}, \citenamefont
  {Ardelt}, \citenamefont {Ihn}, \citenamefont {Ensslin}, \citenamefont
  {Reichl},\ and\ \citenamefont {Wegscheider}}]{rosslernjp2011}%
  \BibitemOpen
  \bibfield  {author} {\bibinfo {author} {\bibfnamefont {C.}~\bibnamefont
  {Rossler}}, \bibinfo {author} {\bibfnamefont {S.}~\bibnamefont {Baer}},
  \bibinfo {author} {\bibfnamefont {E.}~\bibnamefont {de~Wiljes}}, \bibinfo
  {author} {\bibfnamefont {P.-L.}\ \bibnamefont {Ardelt}}, \bibinfo {author}
  {\bibfnamefont {T.}~\bibnamefont {Ihn}}, \bibinfo {author} {\bibfnamefont
  {K.}~\bibnamefont {Ensslin}}, \bibinfo {author} {\bibfnamefont
  {C.}~\bibnamefont {Reichl}}, \ and\ \bibinfo {author} {\bibfnamefont
  {W.}~\bibnamefont {Wegscheider}},\ }\href
  {http://stacks.iop.org/1367-2630/13/i=11/a=113006} {\bibfield  {journal}
  {\bibinfo  {journal} {New Journal of Physics}\ }\textbf {\bibinfo {volume}
  {13}},\ \bibinfo {pages} {113006} (\bibinfo {year} {2011})}\BibitemShut
  {NoStop}%
\end{thebibliography}

%

\end{document}